\documentclass[aps,prd,preprint,nofootinbib,groupedaddress]{revtex4-1}
\usepackage[utf8]{inputenc}
\usepackage{amsmath,mathrsfs,amssymb}
\usepackage{braket}
\usepackage{fancyhdr}
\usepackage{hyperref,graphicx}
\usepackage[stable]{footmisc}
\usepackage{comment}
\usepackage{mathtools}
\usepackage{here}
\usepackage[top=30truemm, bottom=30truemm, left=25truemm, right=25truemm]{geometry}

\begin{document}
\title{Partner formula for an arbitrary moving mirror in $1+1$ dimensions}
\author{Takeshi Tomitsuka}
\author{Koji Yamaguchi}
\author{Masahiro Hotta}
\affiliation{Graduate School of Science, Tohoku University, Sendai 980-8578 Japan}

\begin{abstract}
In the information loss problem in black hole evaporation, the investigation of
the purification partner of a Hawking particle is crucial. It is a well-known fact that the 3+1
dimensional spherically symmetric gravitational collapse can be
mimicked by 1+1 dimensional moving mirror models. 
Since a detected particle in field theory is defined by what a particle detector observes, the diversity of detector designs yields a variety of particles and their partners. We provide a formula of generalized partners of detected particles emitted out of mirrors in an arbitrary
motion for any Gaussian state in a free massless scalar field theory. Using our
formula, we directly demonstrate information storage about initial phase information
in a pure state of a detected particle and its partner. The form of the partner drastically changes depending on the detailed designs of particle detectors for 
Hawking radiation. In the case of a detected particle and its partner sensitive to information about initial phase, spatial configurations of the partner
has long tails in a stage where only zero-point fluctuation is emitted out of the mirror.
\end{abstract}

\maketitle
\section{Introduction}
The information loss problem in black hole evaporation was posed first by
Hawking in 1976 \cite{H}. Suppose that a black hole is formed by a collapsing star. 
The black hole evaporates by emitting the Hawking radiation with a thermal spectrum, which does not depend on the details of the initial phase of black hole formation.
Therefore, it seems that some information about the initial phase cannot be reconstructed only from the Hawking radiation, and cannot be lost during the black hole evaporation process. This problem is a longstanding issue of fundamental
physics during more than four decades. From the viewpoint of quantum
information theory, the problem can be recast into a quest of a purification
partner of the Hawking radiation if quantum gravity maintains unitarity. Since
quantum gravity theory has not been completed yet, nobody knows the correct
partner. Many possible candidates have been proposed including baby
universe\ \cite{D}\cite{Z}, massive remnant\cite{A}\cite{B}\cite{G}, and
the Hawking radiation itself proposed by Page \cite{P}, zero-point fluctuation of
quantum fields \cite{W}\cite{HSU}\cite{HS}, and soft hairs \cite{HSS}
\cite{HPS}.

Just like in the Page scenario, we assume that quantum gravity effects become negligibly small in null future infinity. Hawking particles in the radiation of free
fields should have partners with degrees of freedom of the same fields. The composite system of a Hawking particle and its partner is in a pure state and carries the
initial phase information of the gravitational collapse. The 
partner quest in realistic 3+1 dimensional cases encounters
difficulties of unknown quantum gravity dynamics and cannot be achieved.

In this paper, we adopt a detour route. Similar to black holes, quantum fields scattered by some accelerated mirrors have a thermal spectrum \cite{MM}\cite{F}\cite{M2}\cite{CW}. The radiation can be approximately interpreted as the Hawking radiation emitted by
spherically symmetric black hole formation in 3+1 dimensions \cite{W}. The
mirror trajectory shape reflects a part of the initial parameters of
a 3+1 dimensional gravitational collapse. The emitted Hawking particle is
observed by an Unruh-De Witt detector \cite{Unruh}\cite{Dewitt}. Note that a
wide variety of detector designs are possible even if we use a localized detector
. Since a particle in field theory is defined by what a particle detector observes \cite{Unruh}, the diversity of detector designs yields a variety of particles and their partners \cite{TYH}\cite{TYH2}. The information of the mirror trajectory is stored in
each pair of the detected particle and its partner. We provide a formula of
generalized partners in an arbitrary mirror motion for any Gaussian state in a free massless scalar field. Using this formula, we directly demonstrate information storage about initial phase information. 
The spatial form of the partner is specified by its weighting functions introduced in Sec. \ref{section_3}. 
We find two distinct cases of the partners. In one case, the weighting functions of the detector modes are well localized in the Hawking radiation stage. The weighting functions of the partner modes have almost no dependence on the initial phase information. In the other case, the weighting functions of the detector modes have non vanishing tails in the regime of the initial phase. The weighting functions of the partner modes change drastically depending on the initial phase information.

In Sec. \ref{section_2}, we give a short review on a moving mirror model. In Sec. \ref{section_3}, we provide a partner formula for an arbitrary moving mirror in $1+1$ dimension. In Sec. \ref{section_4}, we demonstrate information storage about initial phase information by using our formula. Finally, In Sec. \ref{section_5}, a summary is provided.

In this paper, we adopt natural units, $c=\hbar=k_{B}=1$.

\section{Moving mirror model\label{section_2}}
\begin{figure}[H]
\begin{minipage}{0.47\hsize}
\begin{center}
\includegraphics[width=70mm]{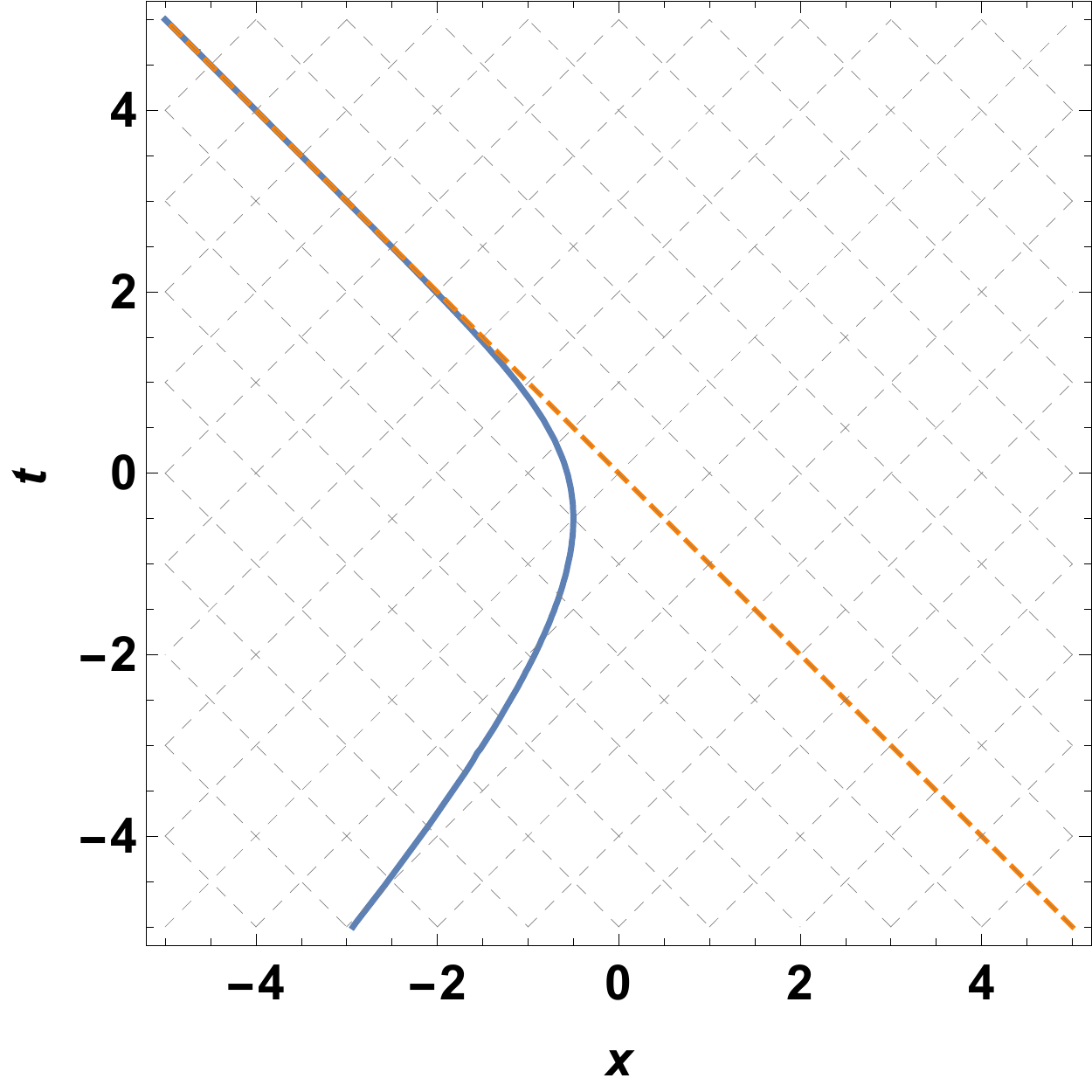}
\caption{The trajectory \eqref{p(u)} $(v_{H}=0)$. $v=0$ (orange dashed line) corresponds to the event horizon.}
\label{p}
\end{center}
\end{minipage}
\hspace{8mm}
\begin{minipage}{0.47\hsize}
\begin{center}
\includegraphics[height=70mm]{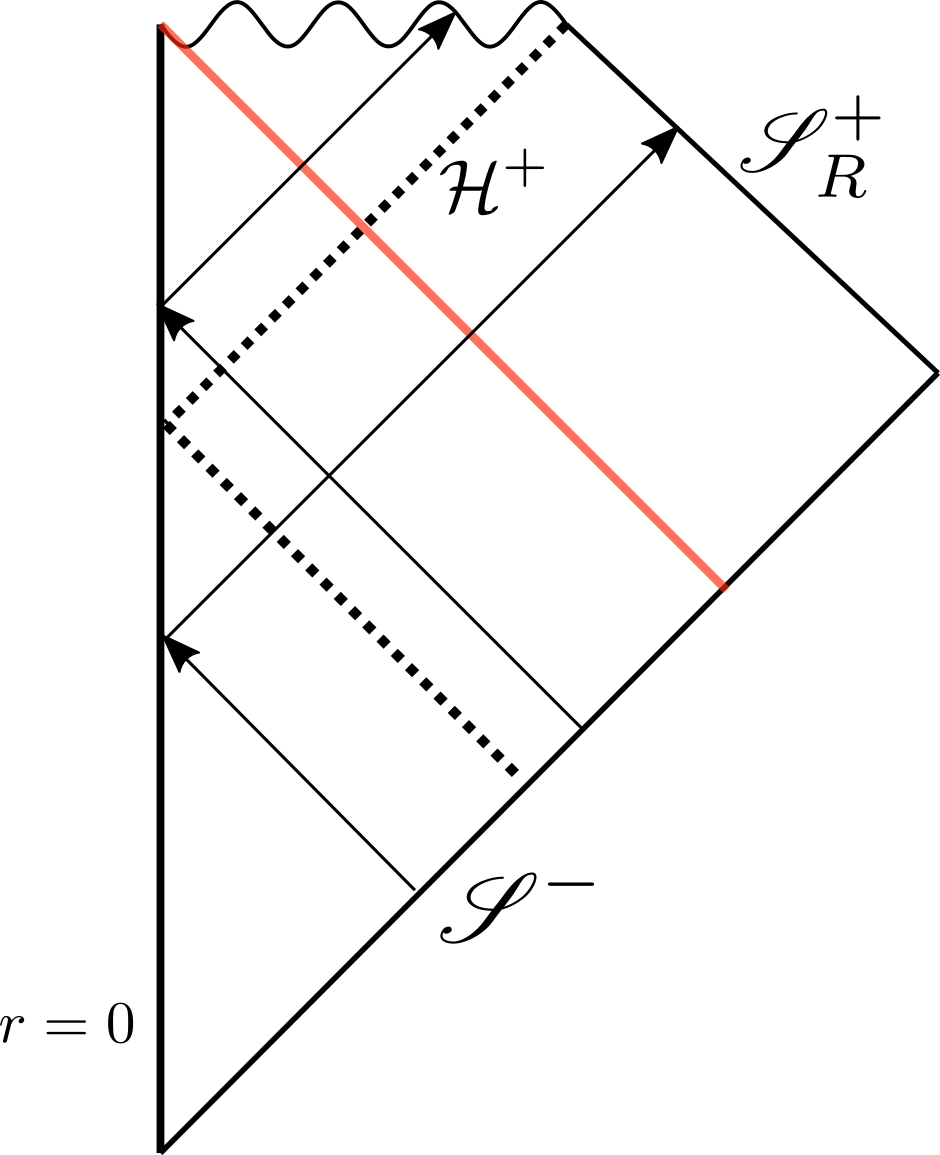}
\caption{Penrose diagram of $3+1$ dimensional spherically symmetric gravitational shell collapse mimics trajectory \eqref{p(u)}. Mirror trajectory corresponds to the origin $(r=0)$. Red line represents collapsing the null shell and dashed line denotes the event horizon.}
\label{Penrose_1}
\end{center}
\end{minipage}
\end{figure}

\begin{figure}[H]
\begin{center}
\includegraphics[width=80mm]{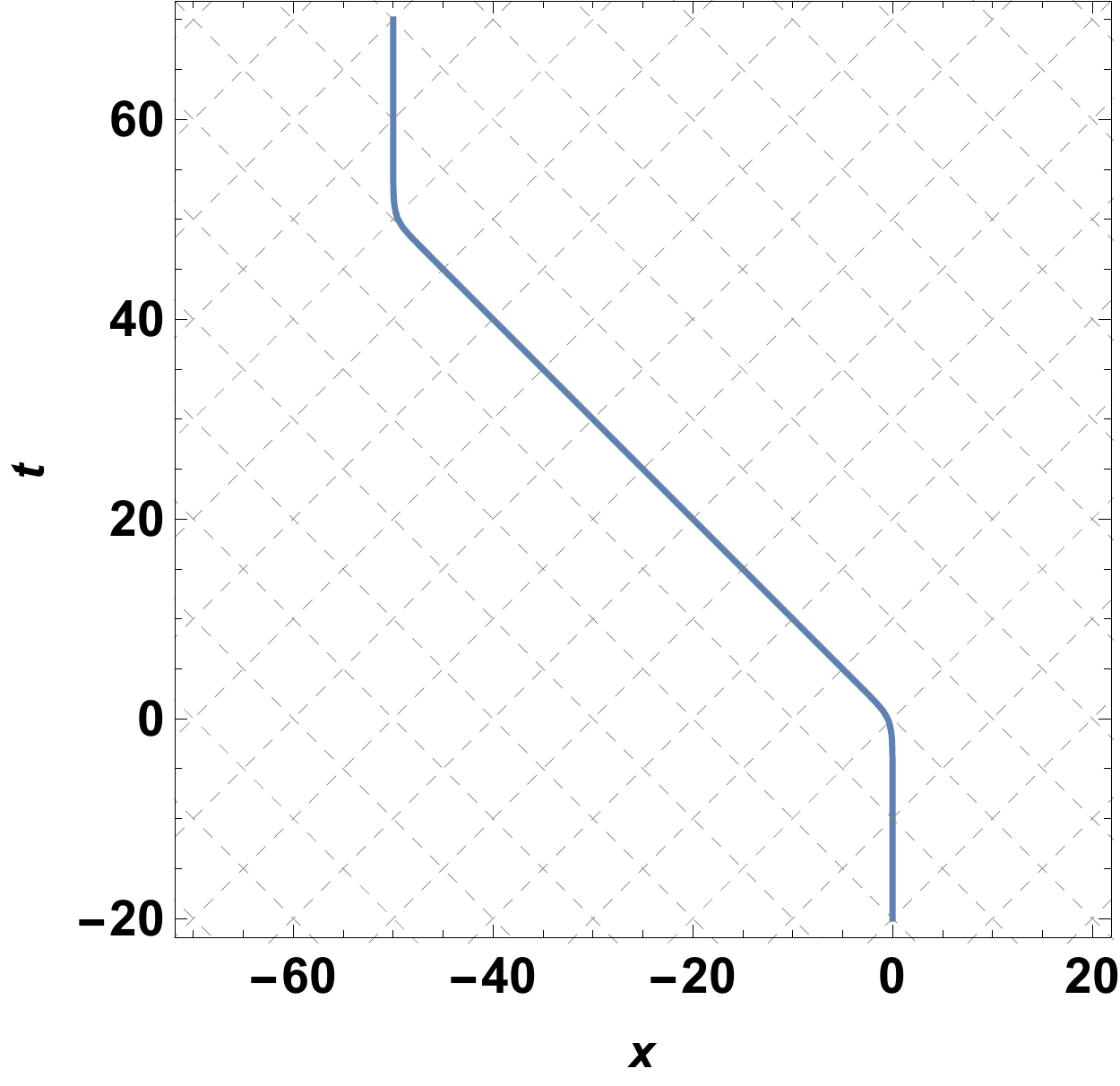}
\caption{The trajectory \eqref{p5(u)} $(\kappa=1,\ h=100)$.}
\label{p5}
\end{center}
\end{figure}

First, we review a moving mirror model which mimics the Hawking radiation and black hole evaporation. 
Consider a flat $1+1$ dimensional spacetime
\begin{equation}
ds^2=-dt^2+dx^2=-dudv,
\end{equation}
where we introduce light-cone coordinates
\begin{equation}
u=t-x\ ,\ v=t+x.
\end{equation}
A massless scaler field $\phi(u,v)$ satisfies the following Klein-Gordon equation
\begin{equation}
\partial_{v}\partial_{u}\phi(u,v)=0
\end{equation}
and vanishes at the location of the mirror $v=p(u)$
\begin{equation}
\label{boundary}
\phi(u,p(u))=0.
\end{equation}
General solutions are written by
\begin{equation}
\phi(u,v)=\phi_{in}(v)+\phi_{out}(u),
\end{equation}
where $\phi_{in}(v)(\phi_{out}(u))$ is an arbitrary function of $v(u)$ and satisfies the following relation
\begin{equation}
\label{boundary_condition}
\phi_{out}(u)=-\phi_{in}(p(u))
\end{equation}
due to Eq.\eqref{boundary}. The incoming field $\phi_{in}(v)$ can be quantized as
\begin{equation}
\hat{\phi}_{in}(v)=\int_{0}^{\infty}\frac{d\omega}{\sqrt{4\pi\omega}}\left(\hat{a}_{\omega}e^{-i\omega v}+\hat{a}_{\omega}^{\dagger}e^{i\omega v}\right),
\end{equation}
where creation and annihilation operators $\hat{a}_{\omega}$ and $\hat{a}_{\omega}^{\dagger}$ satisfy $[\hat{a}_{\omega},\hat{a}_{\omega'}^{\dagger}]=\delta(\omega-\omega')$. The in-vacuum state $\ket{0_{in}}$ is defined by $\hat{a}_{\omega}\ket{0_{in}}=0$. The outgoing field operator $\phi_{out}(u)$ can also be quantized as
\begin{equation}
\hat{\phi}_{out}(u)=\int_{0}^{\infty}\frac{d\omega}{\sqrt{4\pi\omega}}\left(\hat{b}_{\omega}e^{-i\omega u}+\hat{b}_{\omega}^{\dagger}e^{i\omega u}\right),
\end{equation}
where creation and annihilation operators $\hat{b}_{\omega}$ and $\hat{b}_{\omega}^{\dagger}$ satisfy $[\hat{b}_{\omega},\hat{b}_{\omega'}^{\dagger}]=\delta(\omega-\omega')$. The two sets of operator $\{\hat{a}_{\omega},\hat{a}_{\omega}^{\dagger}\}$ and $\{\hat{b}_{\omega},\hat{b}_{\omega}^{\dagger}\}$ are related by
\begin{equation}
\hat{b}_{\omega}=\int_{0}^{\infty}d\omega'\left(\alpha_{\omega\omega'}\hat{a}_{\omega'}+\beta_{\omega\omega'}\hat{a}_{\omega'}^{\dagger}\right),
\end{equation}
where $\alpha_{\omega\omega'}$ and $\beta_{\omega\omega'}$ are Bogoliubov coefficients. If $\beta_{\omega\omega'}\neq 0$, particle creation occurs with a distribution given by
\begin{equation}
\bra{0_{in}}\hat{b}_{\omega}^{\dagger}\hat{b}_{\omega}\ket{0_{in}}=\int_{0}^{\infty}|{\beta_{\omega\omega'}}|^{2}d\omega'
\end{equation}
and the energy flux observed in the future null infinity is given in terms of the mirror trajectory $p(u)$ by
\begin{equation}
\label{energy_flux}
\bra{0_{in}}\hat{T}_{uu}(u)\ket{0_{in}}=-\frac{1}{24\pi}\left[\frac{\partial_{u}^{3}p(u)}{\partial_{u}p(u)}-\frac{3}{2}\left(\frac{\partial_{u}^{2}p(u)}{\partial_{u}p(u)}\right)^{2}\right].
\end{equation}
For example, consider the following trajectory
\begin{equation}
\label{p(u)}
p(u)=v_{H}-\frac{1}{\kappa}e^{-\kappa u}.
\end{equation}
This trajectory mimics an eternal black hole without backreaction of radiation, that is, this black hole never evaporates. $v=v_{H}$ corresponds to the event horizon (Fig.\ref{p}). The incoming field $\hat{\phi}_{in}(v<v_{H})$ is reflected into $\hat{\phi}_{out}(u)$ while $\hat{\phi}_{in}(v \geq v_{H})$ goes through and never comes back. Figure \ref{Penrose_1} represents the Penrose diagram of the corresponding $3+1$ dimensional spherically symmetric gravitational shell collapse. 
The energy flux is constant which can be calculated by Eq.\eqref{energy_flux} as
\begin{equation}
\bra{0_{in}}\hat{T}_{uu}(u)\ket{0_{in}}=\frac{\kappa^{2}}{48\pi},
\end{equation}
with its Hawking temperature $T_{H}=\frac{\kappa}{2\pi}$.
The second example is the following trajectory
\begin{equation}
\label{p5(u)}
p(u)=-\frac{1}{\kappa}\ln{\left(\frac{1+e^{-\kappa u}}{1+e^{\kappa(u-h)}}\right)},
\end{equation}
which mimics a black hole evaporation process with a backreaction of radiation. Parameter $h$ controls the lifetime of the black hole. This mirror is at rest in $u \sim -\infty$, accelerated in $0<u<h$ and finally stopped (Fig.\ref{p5}). Almost constant energy flux $\bra{0_{in}}\hat{T}_{uu}(u)\ket{0_{in}}\sim \frac{\kappa^{2}}{48\pi}$ is emitted during evaporation and no energy flux exists initially and finally. In this case, all incoming fields $\hat{\phi}_{in}(v)$ come back to the future null infinity as $\hat{\phi}_{out}(u)$. 

\section{Partner formula for an arbitrary mirror motion\label{section_3}}
In this section, we provide a formula of generalized partners of detected particles emitted out of mirrors in arbitrary motion for any Gaussian state in a free massless scalar field. In the following, we assume that the trajectory $p(u)$ is a monotonically increasing function and its range is $(-\infty,\infty)$, that is, we consider an evaporation case. 
Let us introduce the incoming momentum operator in the past null infinity ${\mathscr S^{-}}$
\begin{equation}
\hat{\Pi}_{in}(v) \equiv \partial_{v}\hat{\phi}_{in}(v)=-i\int_{0}^{\infty}\sqrt{\frac{\omega}{4\pi}}\left(\hat{a}_{\omega}e^{-i\omega v}-\hat{a}_{\omega}^{\dagger}e^{i\omega v}\right)d\omega.
\end{equation}
We fix the detected particle mode $A$ by a set of canonical operators using $\hat{\Pi}_{in}(v)$
\begin{equation}
(\hat{q}_{A},\hat{p}_{A})=\left(\int q_{A}^{in}(v)\hat{\Pi}_{in}(v)dv,\int  p_{A}^{in}(v)\hat{\Pi}_{in}(v)dv \right),
\end{equation}
where $q_{A}^{in}(v)$ and $p_{A}^{in}(v)$ are real functions of $v$ satisfying 
\begin{equation}
\frac{1}{2}\int_{-\infty}^{\infty}q^{in}_{A}(v)\partial_{v}p_{A}^{in}(v)dv=1
\end{equation}
due to the canonical commutation relation $[\hat{q}_{A},\hat{p}_{A}]=i$ and call them weighting functions. 
We make a local symplectic transformation $S_{A}$
\begin{align}
\left(
\begin{array}{c}
\hat{Q}_{A} \\
\hat{P}_{A} 
\end{array}\right)\equiv
S_{A}\left(
\begin{array}{c}
\hat{q}_{A} \\
\hat{p}_{A}
\end{array}\right)=\left(
\begin{array}{c}
\int_{-\infty}^{\infty}Q_{A}^{in}(v)\hat{\Pi}_{in}(v)dv \\
\int_{-\infty}^{\infty}P_{A}^{in}(v)\hat{\Pi}_{in}(v)dv
\end{array}\right)
\end{align}
such that the operators $(\hat{Q}_{A},\hat{P}_{A})$ are in the standard form (Appendix \ref{symplectic}). Here we defined the functions $(Q_{A}^{in}(v),P_{A}^{in}(v))$ by
\begin{align}
\left(
\begin{array}{c}
Q_{A}^{in}(v) \\
P_{A}^{in}(v)
\end{array}\right)\equiv S_{A}\left(
\begin{array}{c}
q_{A}^{in}(v) \\
p_{A}^{in}(v)
\end{array}\right).
\end{align}
Then, its partner mode B, described as a set of canonical operators $(\hat{Q}_{B},\hat{P}_{B})$, is given by the following: (the derivation is shown in Appendix \ref{derivation_A})
\begin{align}
\label{in_formula_Q}
\hat{Q}_{B}&=\frac{\sqrt{1+g^{2}}}{g}\hat{Q}_{A}+\frac{2}{g}\int_{-\infty}^{\infty}\int_{-\infty}^{\infty}{\rm sgn}(v-v'){\rm Re}\left(\bra{\Psi}\hat{P}_{A}\hat{\Pi}_{in}(v')\ket{\Psi}\right)\hat{\Pi}_{in}(v)dv'dv \\
\label{in_formula_P}
\hat{P}_{B}&=-\frac{\sqrt{1+g^{2}}}{g}\hat{P}_{A}+\frac{2}{g}\int_{-\infty}^{\infty}\int_{-\infty}^{\infty}{\rm sgn}(v-v'){\rm Re}\left(\bra{\Psi}\hat{Q}_{A}\hat{\Pi}_{in}(v')\ket{\Psi}\right)\hat{\Pi}_{in}(v)dv'dv,
\end{align}
where $g\equiv \sqrt{4(\bra{\Psi}\hat{q}_{A}^{2}\ket{\Psi}\bra{\Psi}\hat{p}_{A}^{2}\ket{\Psi}-({\rm Re}\bra{\Psi}\hat{q}_{A}\hat{p}_{A}\ket{\Psi})^{2})-1}$, $\ket{\Psi}$ is an arbitrary Gaussian state, and 
\begin{align}
{\rm sgn}(x)\equiv
\begin{cases}
1 & : x>0 \\
0 & : x=0\\
-1 & : x<0
\end{cases}.
\end{align}
In the future null infinity ${\mathscr S}^{+}$, we define the outgoing momentum operator
\begin{equation}
\hat{\Pi}_{out}(u)\equiv \partial_{u}\hat{\phi}_{out}(u)=-i\int_{0}^{\infty}\sqrt{\frac{\omega}{4\pi}}d\omega(\hat{b}_{\omega}e^{-i\omega u}-\hat{b}_{\omega}^{\dagger}e^{i\omega u}).
\end{equation}
Due to the boundary condition \eqref{boundary_condition}, the relation
\begin{equation}
\label{Pi_relation}
\hat{\Pi}_{out}(u)=-\partial_{u}p(u)\hat{\Pi}_{in}(p(u))
\end{equation}
holds. The operators of mode $A$ can also be written in terms of $\hat{\Pi}_{out}(u)$ 
\begin{align}
\hat{Q}_{A}&=-\int_{-\infty}^{\infty}Q_{A}^{in}(p(u))\hat{\Pi}_{out}(u)du=\int_{-\infty}^{\infty}Q_{A}(u)\hat{\Pi}_{out}(u)du \\
\hat{P}_{A}&=-\int_{-\infty}^{\infty}P_{A}^{in}(p(u))\hat{\Pi}_{out}(u)du=\int_{-\infty}^{\infty}P_{A}(u)\hat{\Pi}_{out}(u)du,
\end{align}
where 
\begin{equation}
Q_{A}(u)\equiv -Q_{A}^{in}(p(u)),\ P_{A}(u)\equiv -P_{A}^{in}(p(u)).
\end{equation} 
These are the weighting functions in the future null infinity ${\mathscr S}^{+}$.
We can get the formula in terms of $\hat{\Pi}_{out}(u)$ from Eqs.\eqref{in_formula_Q},\eqref{in_formula_P} using $v=p(u)$ and the relation \eqref{Pi_relation} 
\begin{align}
\label{out_formula_Q}
\hat{Q}_{B}&=\frac{\sqrt{1+g^{2}}}{g}\hat{Q}_{A}+\frac{2}{g}\int_{-\infty}^{\infty}\int_{-\infty}^{\infty}{\rm sgn}\left(p(u)-p(u')\right){\rm Re}\left(\bra{\Psi}\hat{P}_{A}\hat{\Pi}_{out}(u')\ket{\Psi}\right)\hat{\Pi}_{out}(u)du'du \\
\label{out_formula_P}
\hat{P}_{B}&=-\frac{\sqrt{1+g^{2}}}{g}\hat{P}_{A}+\frac{2}{g}\int_{-\infty}^{\infty}\int_{-\infty}^{\infty}{\rm sgn}\left(p(u)-p(u')\right){\rm Re}\left(\bra{\Psi}\hat{Q}_{A}\hat{\Pi}_{out}(u')\ket{\Psi}\right)\hat{\Pi}_{out}(u)du'du.
\end{align}
This is the partner formula for an arbitrary moving mirror in $1+1$ dimension and the main result in this paper.
Entanglement entropy between the mode $A$ and $B$ is given by
\begin{equation}
S(g)=\sqrt{1+g^{2}}\ln\left(\frac{\sqrt{1+g^{2}}+1}{g}\right)+\ln\left(\frac{g}{2}\right).
\end{equation}
For the vacuum Gaussian state $\ket{\Psi}=\ket{0_{in}}$, this formula is more simplified. The weighting functions are given by
\begin{align}
\label{out_formula_Qw}
Q_{B}(u)&=\frac{\sqrt{1+g^2}}{g}Q_{A}(u)+\frac{1}{g}\int_{-\infty}^{\infty}\Delta(p(u)-p(u'))P_{A}(u')\partial_{u'}p(u')du' \\
\label{out_formula_Pw}
P_{B}(u)&=-\frac{\sqrt{1+g^2}}{g}P_{A}(u)+\frac{1}{g}\int_{-\infty}^{\infty}\Delta(p(u)-p(u'))Q_{A}(u')\partial_{u'}p(u')du',
\end{align}
where we have defined
\begin{equation}
\Delta(x-x')\equiv \frac{i}{2\pi}\int_{-\infty}^{\infty}d\omega\ {\rm sgn}(\omega)e^{-i\omega(x-x')}.
\end{equation}
(the derivation is shown in Appendix \ref{derivation_B})
\section{Application to the simple evaporation mirror model\label{section_4}}
\begin{figure}[H]
\begin{minipage}{0.45\hsize}
	\begin{center}
	 \includegraphics[width=70mm]{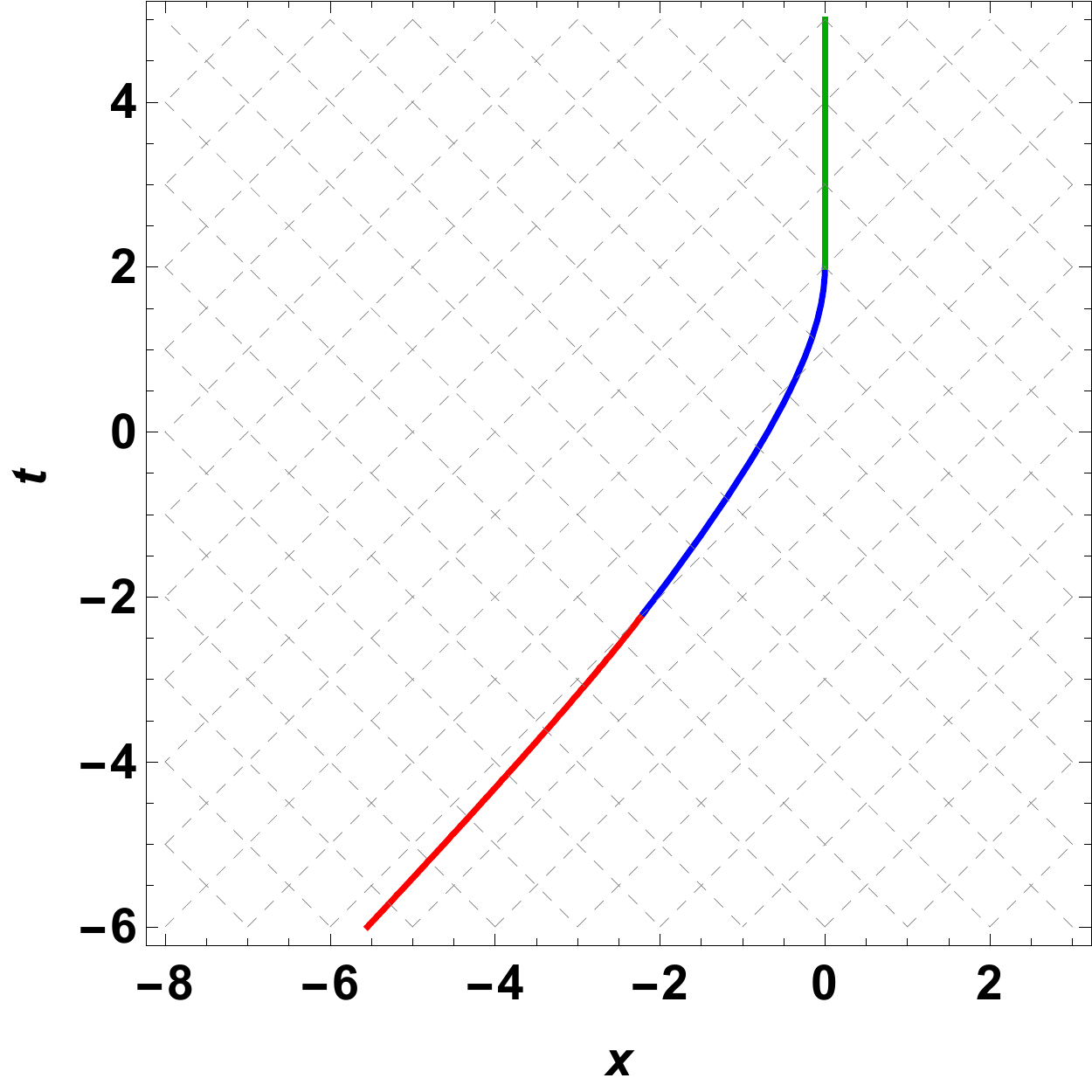}
	 \caption{Mirror trajectory \eqref{p4(u)} $(\kappa_{1}=3,\ \kappa_{2}=1,\ h=2)$. Red line represents the initial phase, blue line represents the  Hawking radiation stage and green line represents the no radiation stage. }
	 \label{p4}
	\end{center}
\end{minipage}
\hspace{8mm}
\begin{minipage}{0.45\hsize}
	\begin{center}
	\includegraphics[width=70mm]{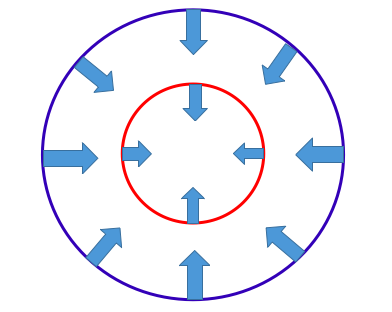}
	\caption{Two-step spherically symmetric gravitational shell collapse. First shell collapse (red circle) is described by red line in Fig.\ref{p4} and second one (blue circle) is described  by blue line in Fig.\ref{p4}.}
	\label{collapse}
	\end{center}
\end{minipage}
\end{figure}

\begin{figure}[H]
 \begin{center}
  \includegraphics[width=80mm]{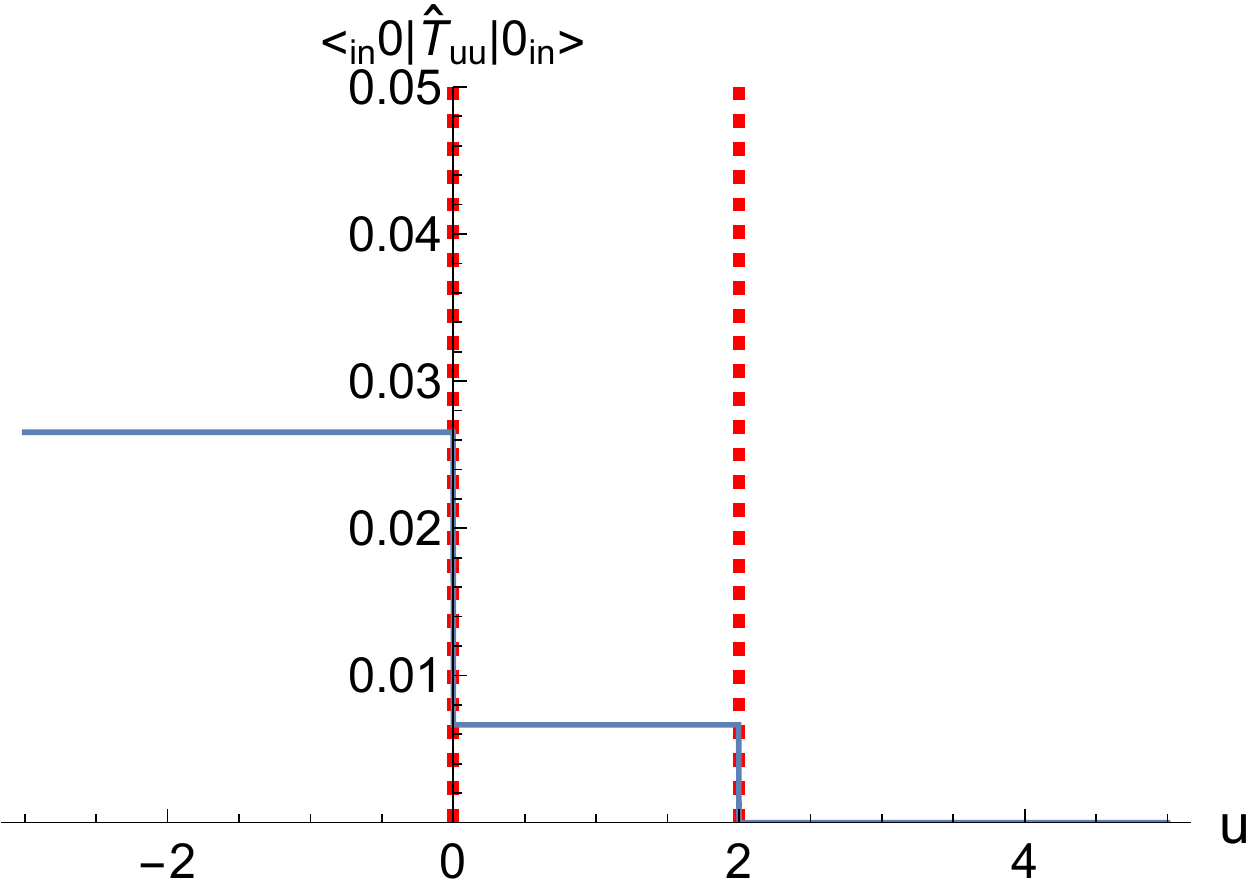}
  \caption{Energy flux $\bra{0_{in}}\hat{T}_{uu}(u)\ket{0_{in}}$ for the trajectory \eqref{p4} $(\kappa_{1}=3,\ \kappa_{2}=1,\ h=2)$. In the initial phase $(u \leq 0)$, the Hawking temperature $T_{1}=\frac{3}{2\pi}$, in the Hawking radiation stage $(0 < u \leq 2)$, $T_{2}=\frac{1}{2\pi}$. Red dashed lines represent energy fluxes proportional to delta function at the junctions of the trajectory. Blue solid line represents energy flux without their contributions.}
   \label{T4}
 \end{center}
\end{figure}

\begin{figure}[H]
 \begin{minipage}{0.45\hsize}
  \begin{center}
   \includegraphics[width=70mm]{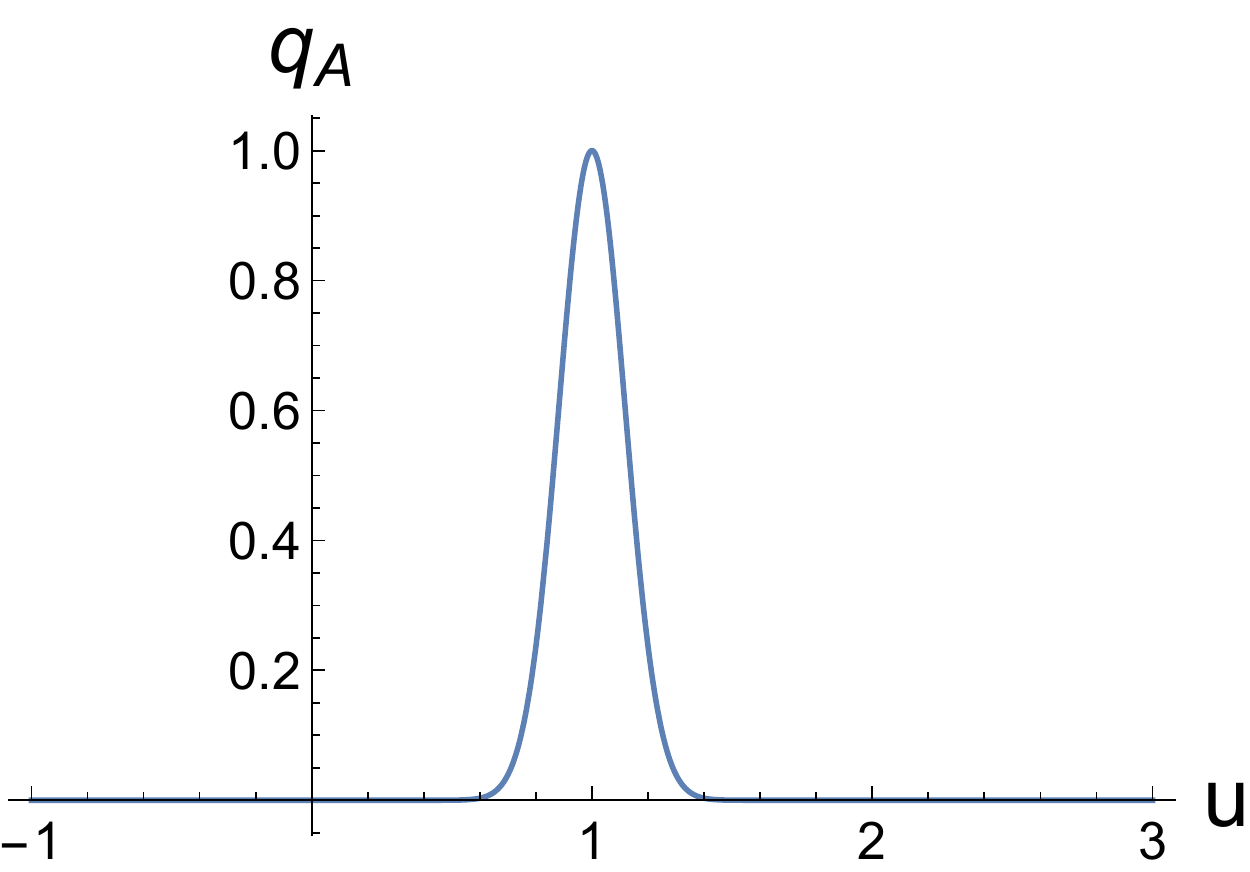}
  \end{center}
  \caption{The weighting function of mode $A$ $q_{A}(u)$ $(C=6)$ well localized in the Hawking radiation stage $(0<u\leq 2)$.}
  \label{qA}
 \end{minipage}
 \hspace{3mm}
 \begin{minipage}{0.45\hsize}
  \begin{center}
   \includegraphics[width=70mm]{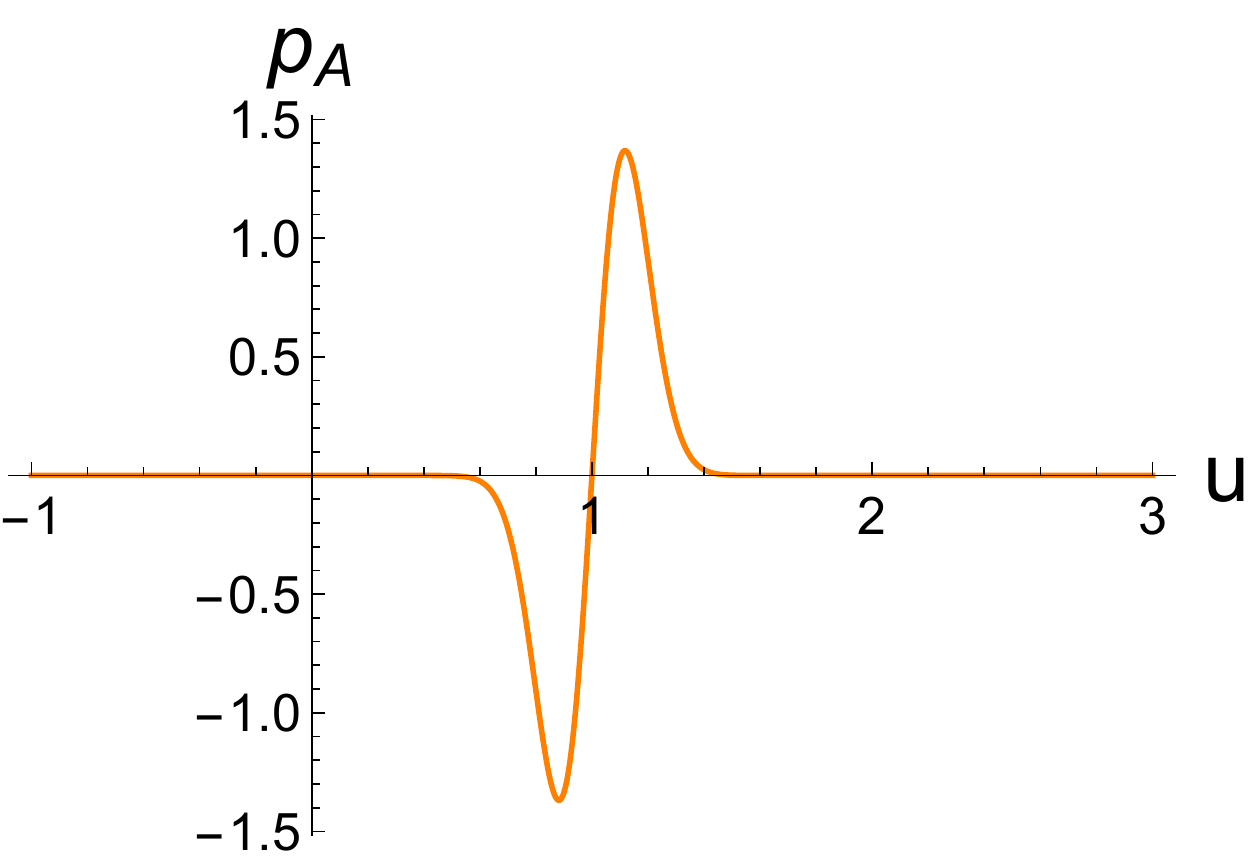}
  \end{center}
  \caption{The weighting function of mode $A$ $p_{A}(u)$ $(C=6)$ well-localized in the  Hawking radiation stage $(0<u\leq 2)$.}
  \label{pA}
 \end{minipage}
\end{figure}

\begin{figure}[H]
\begin{center}
\includegraphics[width=70mm]{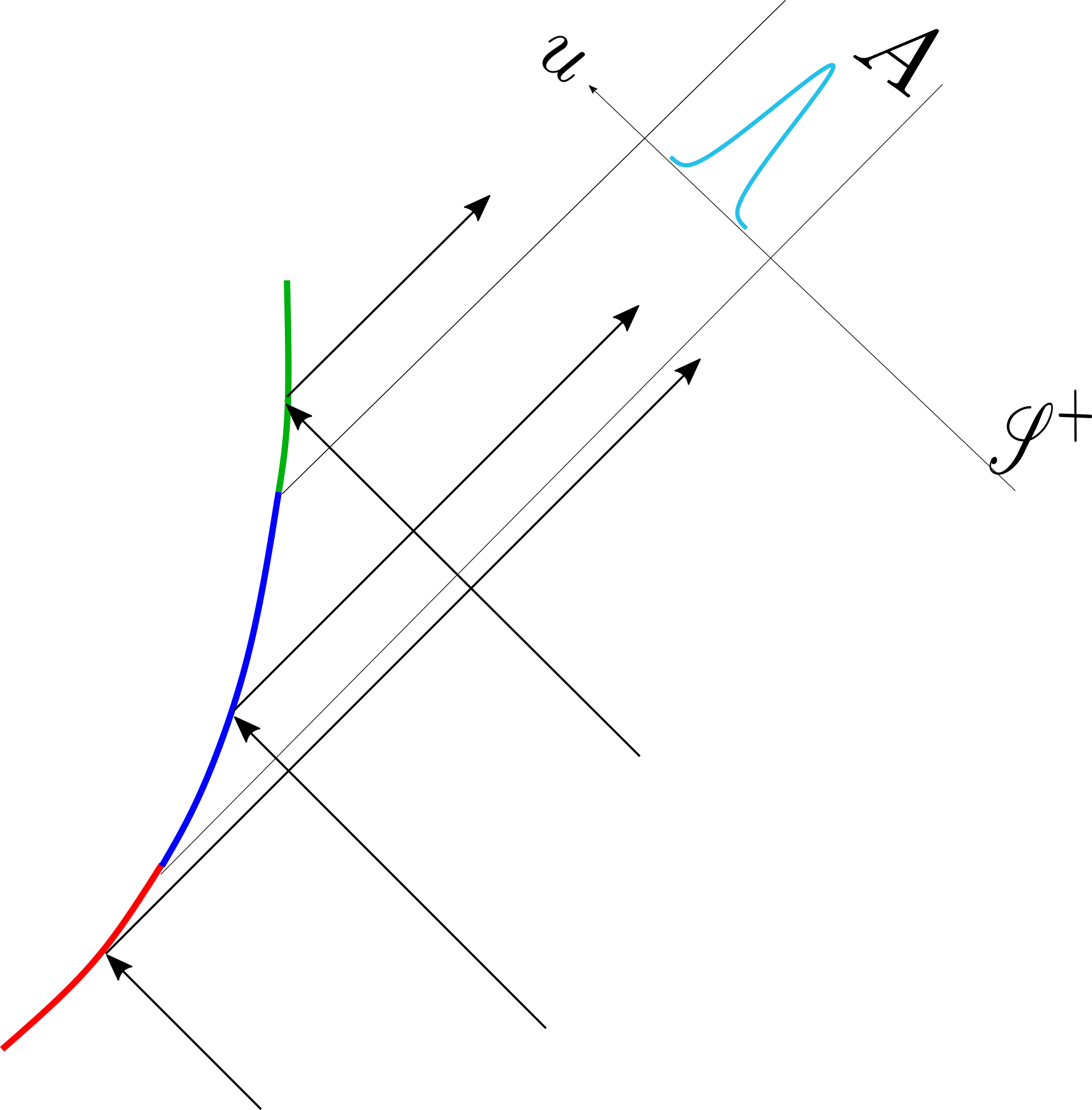}
\caption{The schematic picture of this setup. The weighting functions of mode $A$ are well localized in the Hawking radiation stage.}
\label{mirror_1}
\end{center}
\end{figure}

We want to investigate how the initial phase information of a moving mirror is stored in a pure state of a pair of detected particle and its partner. To do this, we consider the following simple trajectory
\begin{align}
\label{p4(u)}
p(u)=
\begin{cases}
-\frac{e^{\kappa_{2}h}}{\kappa_{1}}e^{-\kappa_{1}u}+\left(\frac{1}{\kappa_{1}}-\frac{1}{\kappa_{2}}\right)e^{\kappa_{2}h}+h+\frac{1}{\kappa_{2}} & (u\leq 0) \\
-\frac{e^{\kappa_{2}h}}{\kappa_{2}}e^{-\kappa_{2}u}+h+\frac{1}{\kappa_{2}} & (0 < u \leq h) \\
\ \ u & (h<u)
\end{cases}
\end{align}
which has three stages, namely, the initial phase $(u\leq 0)$, the Hawking radiation stage $(0<u\leq h)$ and the no radiation stage $(h < u)$. Figure \ref{p4} shows the trajectory \eqref{p4(u)}. Note that dynamics of realistic black hole formation is different from that of the moving mirror. Thus, the causal structure of the black hole case cannot be correctly described by the moving mirror model.
However, there exists an apparent information loss in the moving mirror model since it seems that Hawking radiation does not depend on the initial phase information, as already stressed above. 
It is significant to address the question where the information is stored in the final quantum states.
This trajectory mimics the two-step spherically symmetric gravitational shell collapse and evaporation process (Fig.\ref{collapse}). The initial phase corresponds to the first shell collapse, the Hawking radiation stage to the second shell collapse, and the no radiation stage to the region where only zero-point fluctuation exists, respectively. 
Figure \ref{T4} shows the energy flux for the trajectory\eqref{p4(u)}. The Hawking temperature in the initial phase $(0\leq u)$ and in the Hawking radiation stage $(0 < u \leq h)$ are $T_{1}=\frac{\kappa_{1}}{2\pi}$ and $T_{2}=\frac{\kappa_{2}}{2\pi}$, respectively. 
In this setup, we assume that $\ket{\Psi}=\ket{0_{in}}$ in the past null infinity ${\mathscr S}^{-}$ and it evolves under the trajectory \eqref{p4(u)}.
Fixing parameters $\kappa_{2}=1,\ h=2$ and setting the detected particle mode $A$ in the future null infinity ${\mathscr S}^{+}$,  we investigate how its partner mode $B$ changes with respect to $\kappa_{1}$ which is initial phase information. Let us set the mode $A$ by the following form: 
\begin{align}
\hat{q}_{A}&=\int_{-\infty}^{\infty}q_{A}(u)\hat{\Pi}_{out}(u)du\\
\hat{p}_{A}&=\int_{-\infty}^{\infty}p_{A}(u)\hat{\Pi}_{out}(u)du \\
q_{A}(u)&=e^{-C^{2}(u-\frac{h}{2})^{2}} \\
p_{A}(u)&=4C\sqrt{\frac{2}{\pi}}\left(u-\frac{h}{2}\right)e^{-C^{2}(u-\frac{h}{2})^{2}}
\end{align}
satisfying $\frac{1}{2}\int_{-\infty}^{\infty}q_{A}(u)\partial_{u}p_{A}(u)du=1$ (which corresponds to $[\hat{q}_{A},\hat{p}_{A}]=i$). A positive parameter $C$ controls the degree of localization in the Hawking radiation stage. 
First, we set them well-localized in the Hawking radiation stage $(C=6)$ (Fig.\ref{qA},\ref{pA}). The schematic picture of this setup is shown in Fig.\ref{mirror_1}.

\begin{figure}[H]
 \begin{minipage}{0.47\hsize}
  \begin{center}
   \includegraphics[width=70mm]{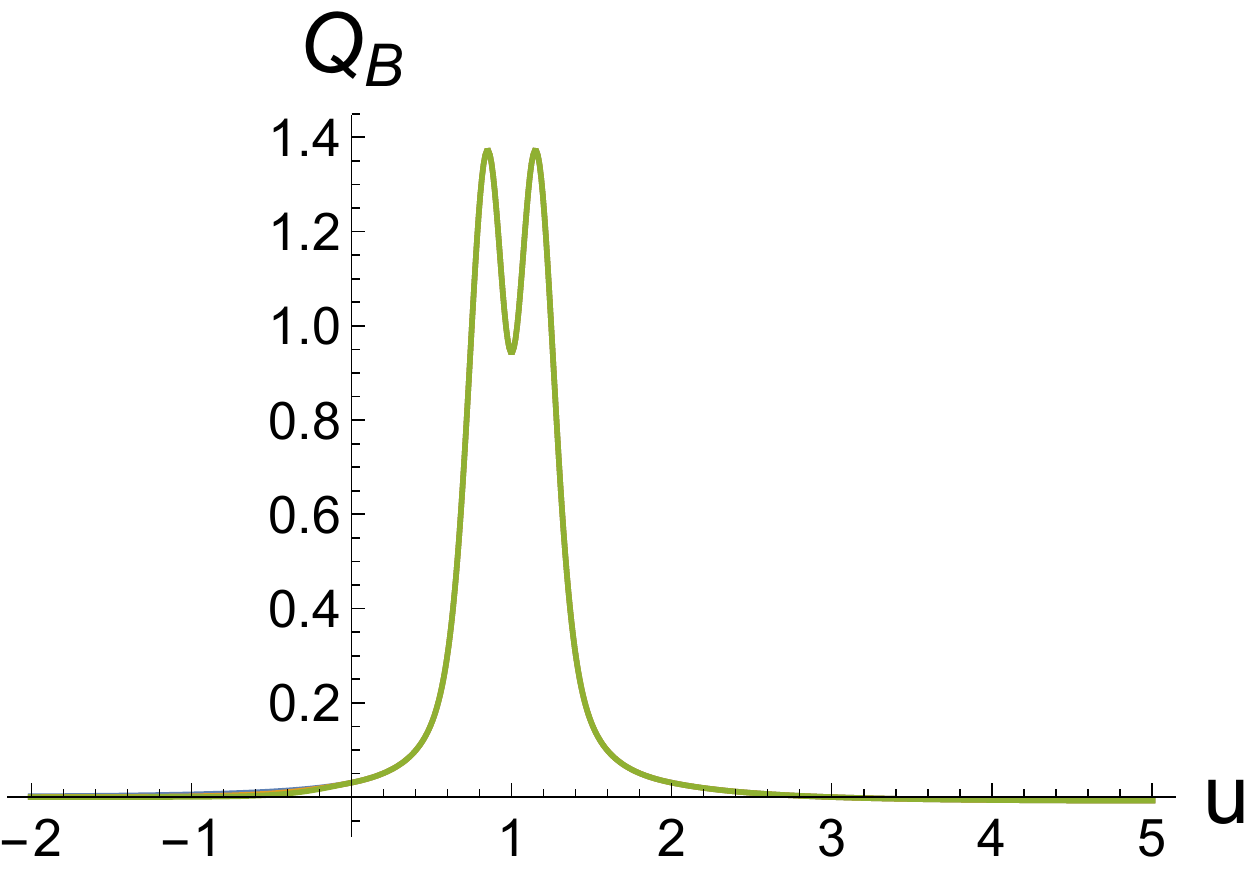}
  \end{center}
  \caption{The weighting function of partner mode $B$ $Q_{B}(u)$ for the localized mode $A$.}
  \label{QB}
 \end{minipage}
 \hspace{3mm}
 \begin{minipage}{0.47\hsize}
  \begin{center}
   \includegraphics[width=70mm]{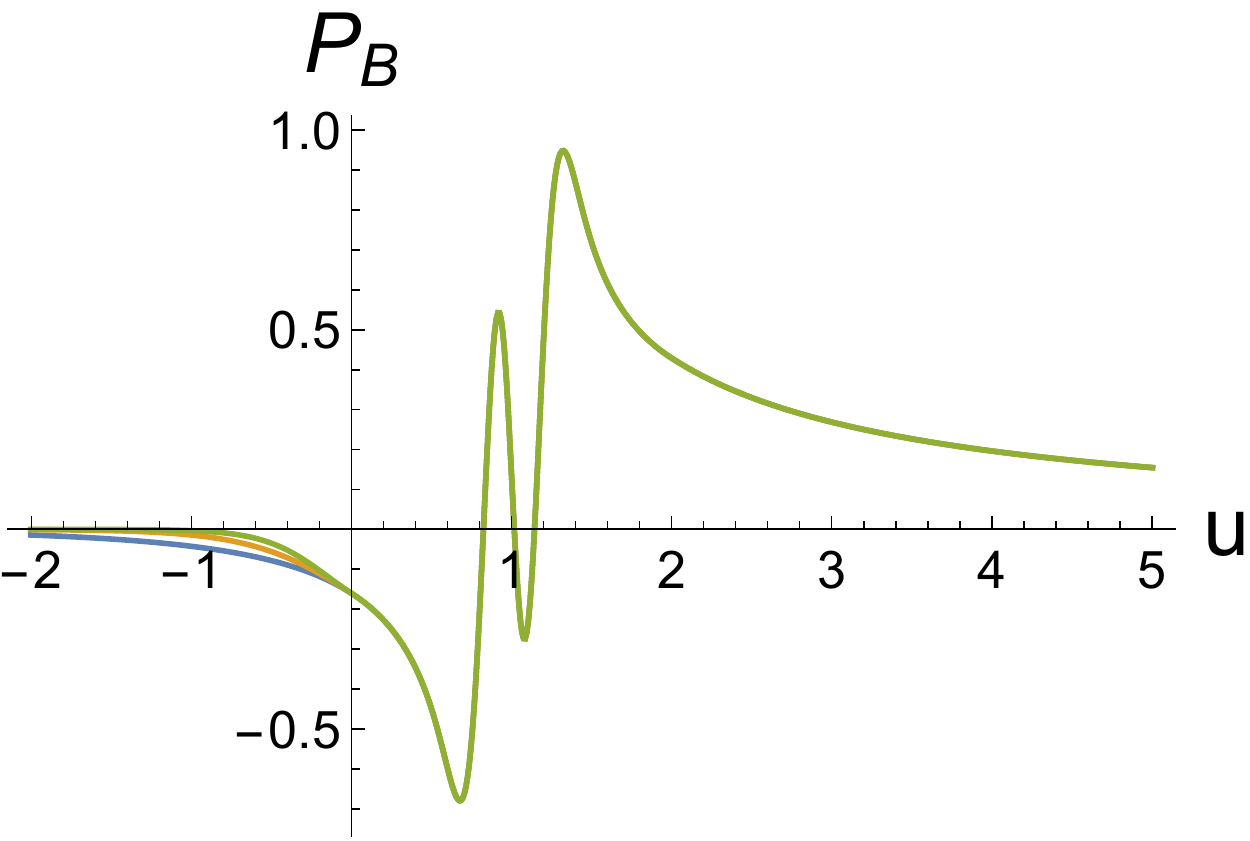}
  \end{center}
  \caption{The weighting function of partner mode $B$ $P_{B}(u)$ for the localized mode $A$.}
  \label{PB}
 \end{minipage}
\end{figure}

\begin{figure}[H]
 \begin{minipage}{0.47\hsize}
  \begin{center}
   \includegraphics[width=70mm]{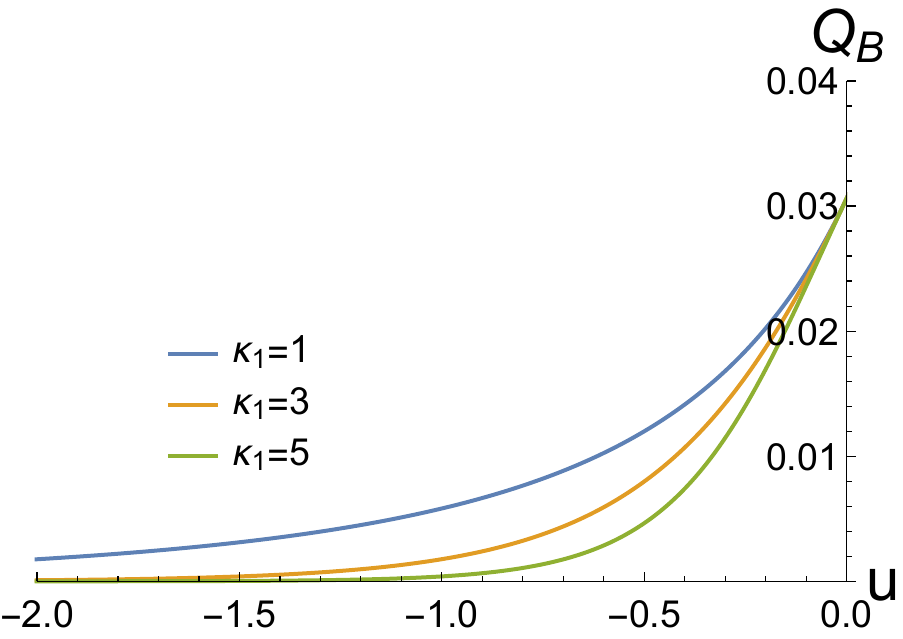}
  \end{center}
  \caption{The weighting function of partner mode $B$ in the initial phase $Q_{B}(u<0)$.}
  \label{QBsub}
 \end{minipage}
 \hspace{3mm}
 \begin{minipage}{0.5\hsize}
  \begin{center}
   \includegraphics[width=70mm]{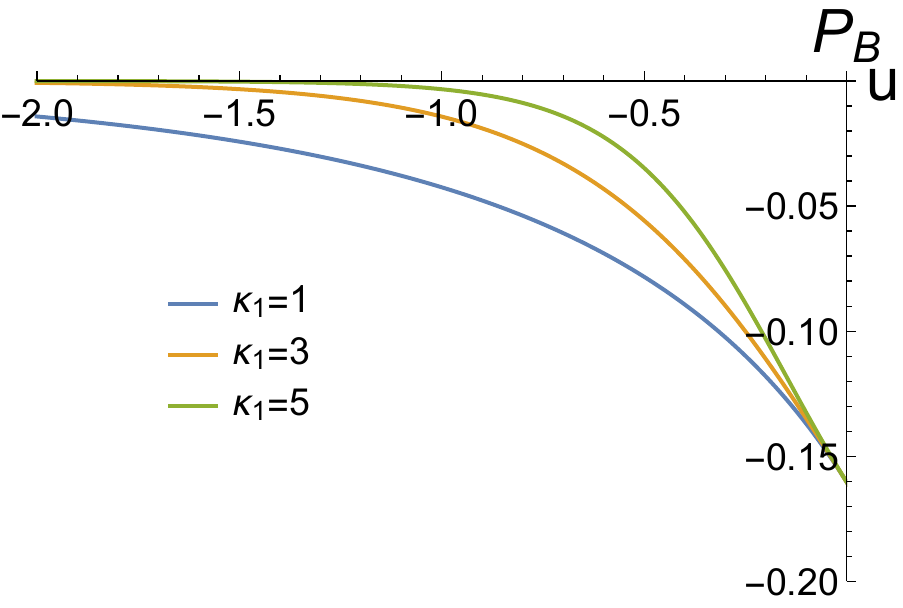}
  \end{center}
  \caption{The weighting function of partner mode $B$ in the initial phase $P_{B}(u<0)$.}
  \label{PBsub}
 \end{minipage}
\end{figure}

For this mode $A$, Figs.\ref{QB} and \ref{PB} show the numerical result of the weighting functions of its partner mode $B$.
Its partner mode $B$ changes little in the initial phase with respect to $\kappa_{1}$ (Figs.\ref{QBsub} and \ref{PBsub}). So we can hardly determine what $\kappa_{1}$ is from this pair of partners, that is, this pair has little initial phase information. The result that there is no change in the thermal and the no radiation stage is clear by the partner formula. When the weighting functions of mode $A$ are well localized in the Hawking radiation stage, the  dependence of $\kappa_{1}$ emerges only in the initial phase of the second terms of the right-hand sides in Eqs.\eqref{out_formula_Qw} and \eqref{out_formula_Pw}. 

\begin{figure}[H]
 \begin{minipage}{0.47\hsize}
  \begin{center}
   \includegraphics[width=70mm]{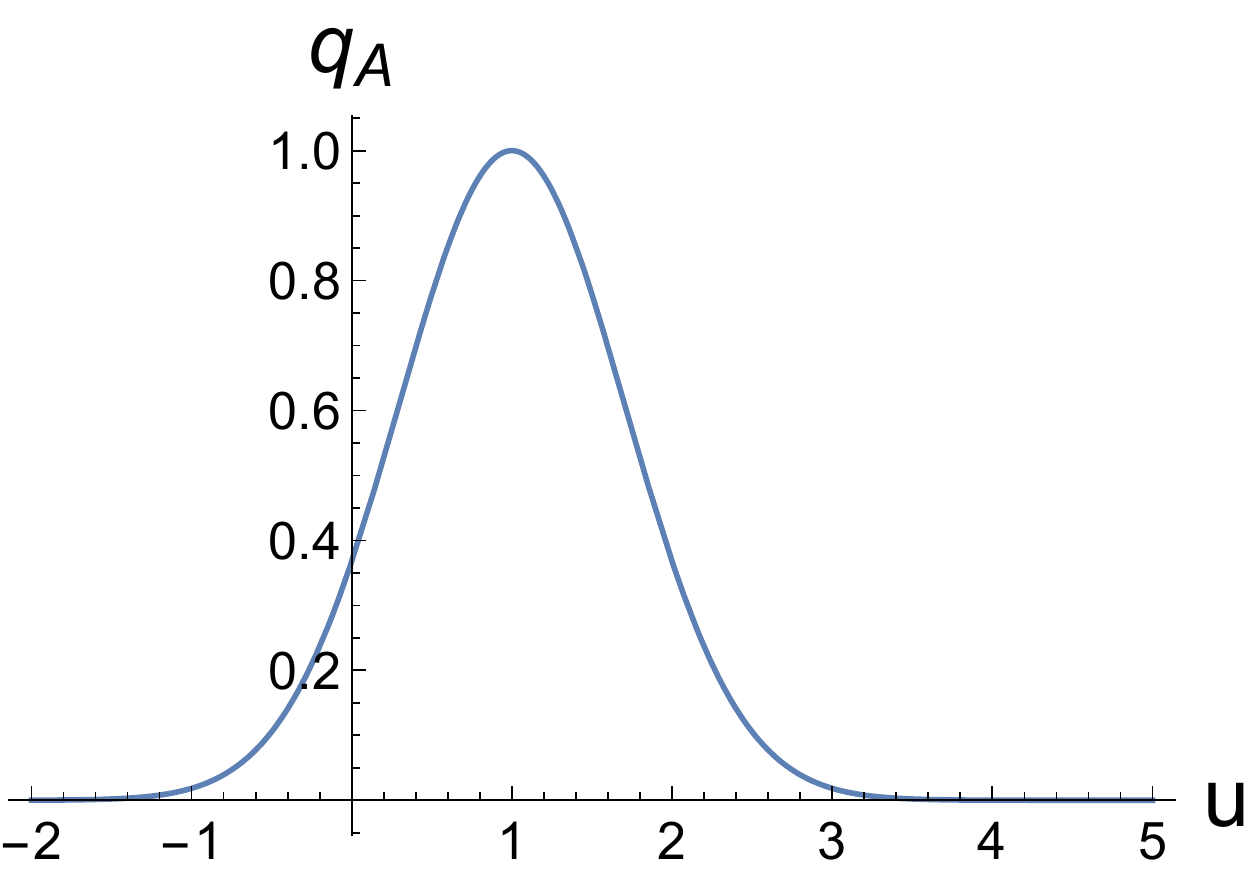}
  \end{center}
  \caption{The weighting function of mode $A$ $q_{A}(u)$ $(C=1)$ nonlocalized in the Hawking radiation stage $(0<u \leq 2)$.}
  \label{qA2}
 \end{minipage}
 \hspace{3mm}
 \begin{minipage}{0.47\hsize}
  \begin{center}
   \includegraphics[width=70mm]{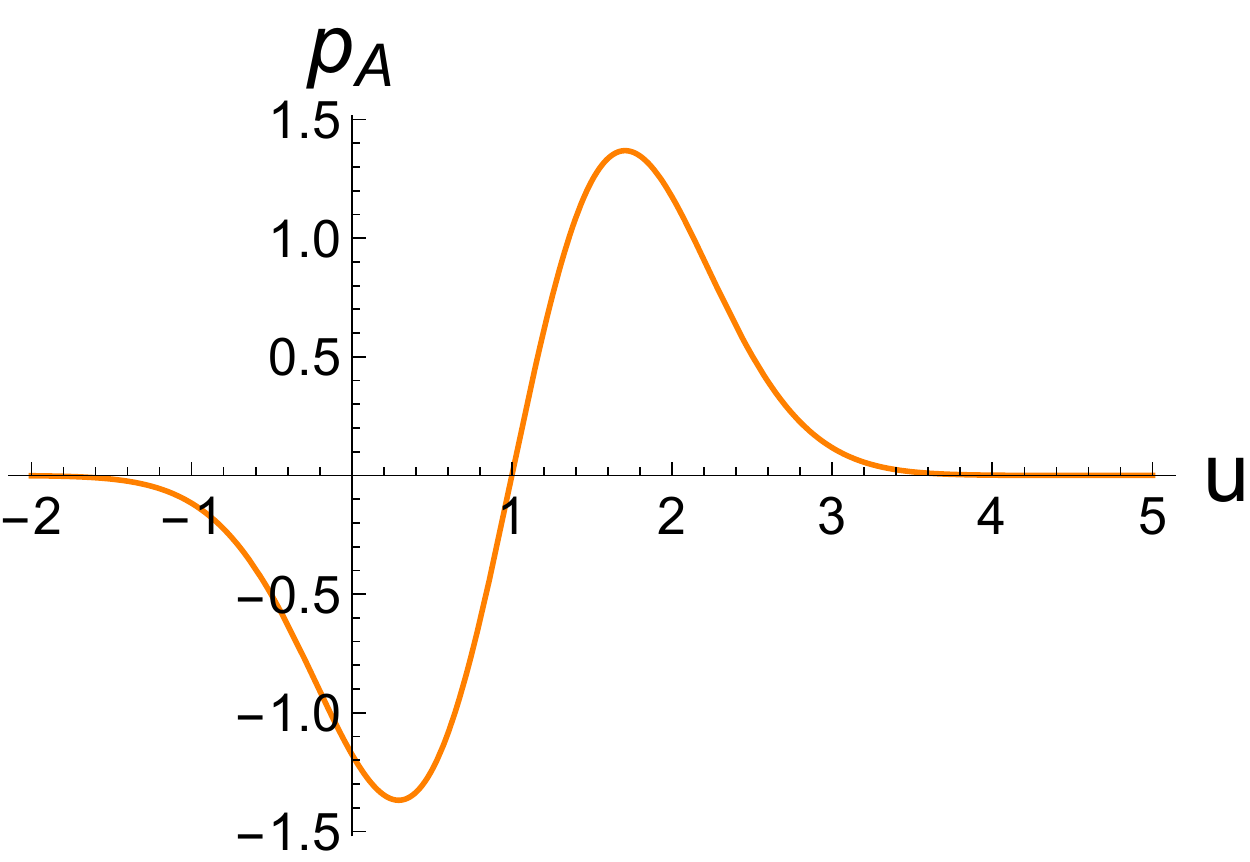}
  \end{center}
  \caption{The weighting function of mode $A$ $p_{A}(u)$ $(C=1)$ nonlocalized in the Hawking stage $(0 < u \leq 2)$.}
  \label{pA2}
 \end{minipage}
\end{figure}

\begin{figure}[H]
\begin{center}
\includegraphics[width=70mm]{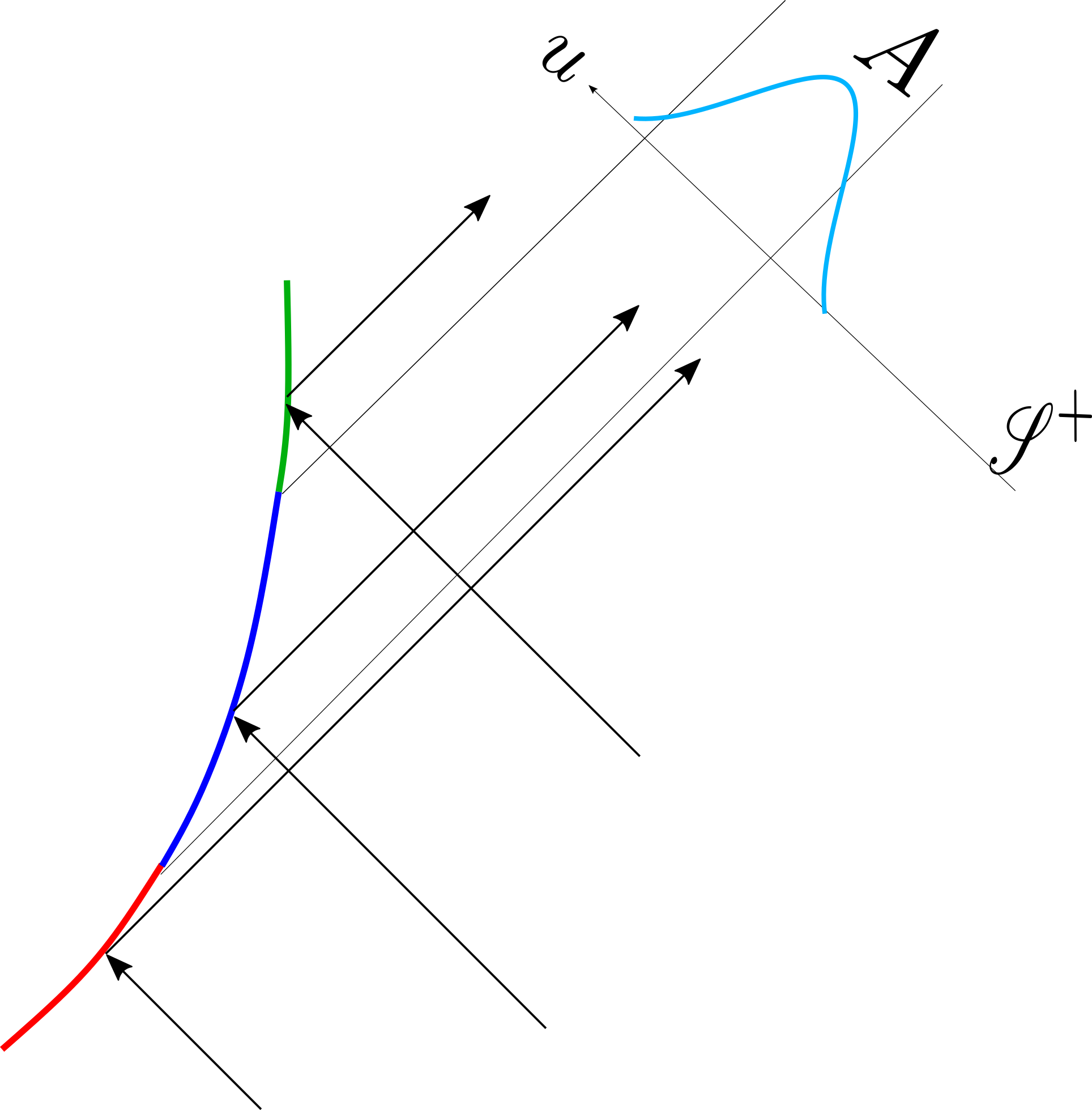}
\caption{The schematic picture of this setup. The weighting functions of mode $A$ are nonlocalized in the Hawking radiation stage which have the nonvanishing tails in the initial phase.}
\label{mirror_2}
\end{center}
\end{figure}

\begin{figure}[H]
 \begin{minipage}{0.47\hsize}
  \begin{center}
   \includegraphics[width=70mm]{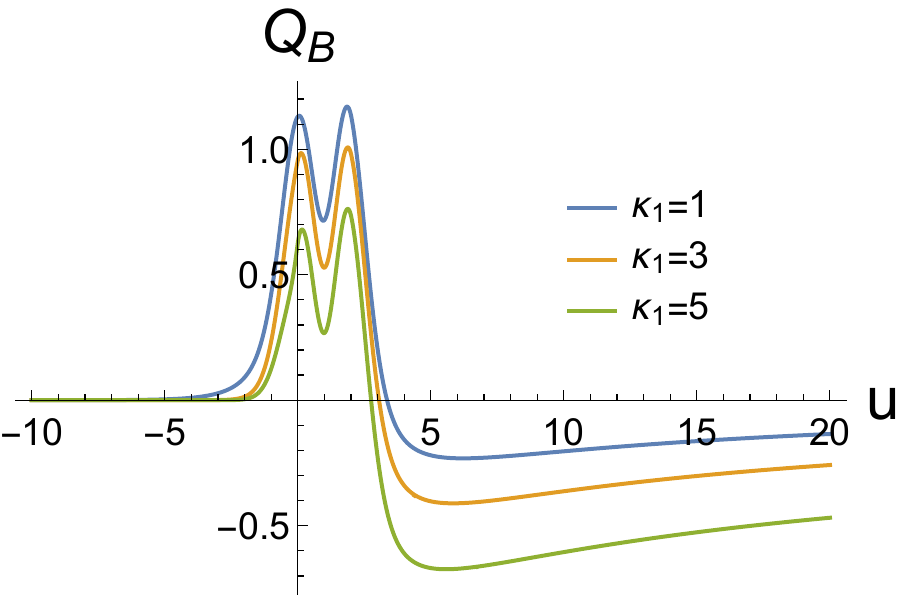}
  \end{center}
  \caption{The weighting function of partner mode $B$ $Q_{B}(u)$ with respect to $\kappa_{1}$ for the  nonlocalized mode $A$.}
  \label{QB2}
 \end{minipage}
 \hspace{3mm}
 \begin{minipage}{0.47\hsize}
  \begin{center}
   \includegraphics[width=70mm]{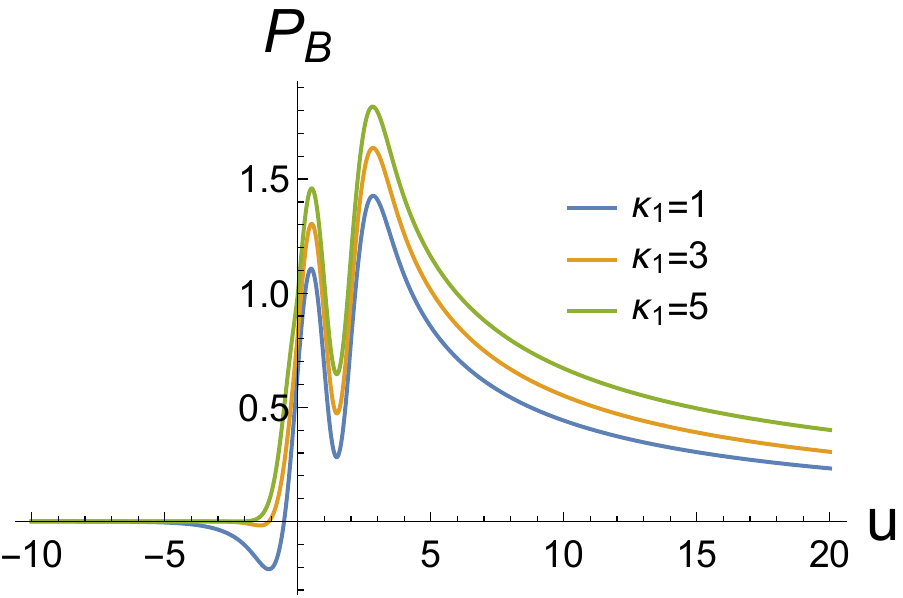}
  \end{center}
  \caption{The weighting function of partner mode $B$ $P_{B}(u)$ with respect to $\kappa_{1}$ for the  nonlocalized mode $A$.}
  \label{PB2}
 \end{minipage}
\end{figure}

Next, we let the weighting functions of mode $A$ nonlocalized a little $(C=1)$ which have nonvanishing tails in the initial phase, as depicted in 
Figs.\ref{qA2} and \ref{pA2}. The schematic picture of this setup is shown in Fig.\ref{mirror_2}. 
For this mode $A$, the weighting functions of its partner mode $B$ are shown in Figs.\ref{QB2} and  \ref{PB2}.
 Its partner mode $B$ changes dramatically with respect to $\kappa_{1}$ in all stages due to the second terms of the right-hand sides in Eqs.\eqref{out_formula_Qw} and \eqref{out_formula_Pw}. Then we can determine what $\kappa_{1}$ is from this pair of partners, that is, this pair has the initial phase information. Note that the partner mode $B$ has a long tail in the no radiation stage $(u>2)$, so initial phase information is also stored in the correlation between the mode $A$ and zero-point fluctuation, pointed out in \cite{HSU}.
\section{Summary\label{section_5}}
For an arbitrary Gaussian state, we provide a formula of generalized partners for an arbitrary moving mirror in $1+1$ dimensions [Eqs.\eqref{out_formula_Q},\eqref{out_formula_P}]. Next, we demonstrate information storage about initial phase information in a pair of partners by applying this formula to a simple mirror trajectory \eqref{p4(u)} which mimics two-step spherically symmetric gravitational shell collapse. The sensitivity of a pair of partners to initial phase information depends on the design of the detected particle mode. In particular, for the detected particle mode well-localized in the Hawking radiation stage (Figs.\ref{qA},\ref{pA}), little initial phase information is stored in that pair of partners (Figs.\ref{QB},\ref{PB}). On the other hand, for a little nonlocalized one (Figs.\ref{qA2},\ref{pA2}), initial phase information is stored in it including the correlation with zero-point fluctuation (Figs.\ref{QB2},\ref{PB2}). It would be interesting to extend our formula to the higher dimensional case and two mirror case. The latter could be applicable  in a dynamical Casimir effect experiment where two imperfect reflectional mirrors are used.\\
Recently, Wald wrote an interesting paper \cite{Wald} and proposed a Milne partner of a Hawking particle. Although ordinary entanglement of two particles is spacelike correlation, the entanglement between a Milne particle and a Hawking particle is timelike correlation. Our results in this paper will be useful to explore the above possibility.   

\begin{center}
\section*{Acknowledgment}
\end{center}
We thank the participants in ``Relativistic Quantum Information - North 2019'' at National Cheng Kung University for useful discussion and Michael R.R. Good for useful comments. We appreciate a discussion with Robert M. Wald. This research is partially supported by JSPS KAKENHI Grant Number JP19K03838 (M.H.) and JP18J20057 (K.Y.), and
by the Graduate Program on Physics for the Universe (GP-PU), Tohoku University (T.T and K.Y.).

\appendix
\section{THE LOCAL SYMPLECTIC TRANSFORMATION TO THE STANDARD FORM\label{symplectic}}
By the local symplectic transformation of particle mode A
\begin{align}
\left(
\begin{array}{c}
\hat{Q}_{A} \\
\hat{P}_{A} 
\end{array}
\right)=S_{A}\left(
\begin{array}{c}
\hat{q}_{A} \\
\hat{p}_{A}
\end{array}
\right)=\left(
\begin{array}{cc}
\cos\theta_{A}' & \sin\theta_{A}' \\
-\sin\theta_{A}' & \cos\theta_{A}'
\end{array}\right)\left(
\begin{array}{cc}
e^{\sigma_{A}} & 0 \\
0 & e^{-\sigma_{A}}
\end{array}\right)\left(
\begin{array}{cc}
\cos\theta_{A} & \sin\theta_{A} \\
-\sin\theta_{A} & \cos\theta_{A}
\end{array}\right)\left(
\begin{array}{c}
\hat{q}_{A} \\
\hat{p}_{A}
\end{array}\right)
\end{align}
, we can transform the corresponding canonical operators $(\hat{q}_{A}, \hat{p}_{A})^{T} \rightarrow (\hat{Q}_{A},\hat{P}_{A})^{T}$ such that
\begin{align}
\left(
\begin{array}{cc}
\bra{\Psi}\hat{Q}_{A}^{2}\ket{\Psi} & {\rm Re}\left(\bra{\Psi}\hat{Q}_{A}\hat{P}_{A}\ket{\Psi}\right) \\
{\rm Re}\left(\bra{\Psi}\hat{P}_{A}\hat{Q}_{A}\ket{\Psi}\right) & \bra{\Psi}\hat{P}_{A}^{2}\ket{\Psi}
\end{array}
\right)=\frac{\sqrt{1+g^{2}}}{2}\left(
\begin{array}{cc}
1 & 0 \\
0 & 1
\end{array}
\right),
\end{align}
where $g\equiv \sqrt{4(\bra{\Psi}\hat{q}_{A}^{2}\ket{\Psi}\bra{\Psi}\hat{p}_{A}^{2}\ket{\Psi}-({\rm Re}\bra{\Psi}\hat{q}_{A}\hat{p}_{A}\ket{\Psi})^{2})-1}$, $\ket{\Psi}$ is an arbitrary Gaussian state.
\section{DERIVATION OF EQS.\eqref{in_formula_Q} AND \eqref{in_formula_P} \label{derivation_A}}
We define the canonical operators 
\begin{equation}
\hat{Q}_{\omega}\equiv \frac{1}{\sqrt{2}}(\hat{a}_{\omega}+\hat{a}_{\omega}^{\dagger}),\ \ \ \hat{P}_{\omega}\equiv \frac{1}{\sqrt{2}i}(\hat{a}_{\omega}-\hat{a}_{\omega}^{\dagger})
\end{equation}
and
\begin{equation}
\hat{\boldsymbol{R}}_{\omega}\equiv(\hat{Q}_{\omega},\hat{P}_{\omega})^{T}
\end{equation}
and assume that $\bra{\Psi}\hat{\boldsymbol R}_{\omega}\ket{\Psi}=0$ holds for an arbitrary state $\ket{\Psi}$ without loss of generality by shifting 
\begin{equation}
\hat{Q}_{\omega}\rightarrow \hat{Q}_{\omega}-\bra{\Psi}\hat{Q}_{\omega}\ket{\Psi},\ \hat{P}_{\omega}\rightarrow \hat{P}_{\omega}-\bra{\Psi}\hat{P}_{\omega}\ket{\Psi}.
\end{equation}
Then $\hat{{\boldsymbol R}}_{\omega}$ satisfies
\begin{equation}
[\hat{\boldsymbol R}_{\omega},\hat{\boldsymbol R}_{\omega'}^{T}]=i\Omega(\omega,\omega')
\end{equation}
with
\begin{align}
\Omega(\omega,\omega')\equiv \left(
\begin{array}{cc}
0 & \delta(\omega-\omega') \\
-\delta(\omega-\omega') & 0
\end{array}\right)
\end{align}
The covariance matrix is defined by
\begin{equation}
M(\omega,\omega')\equiv {\rm Re}(\bra{\Psi}\hat{\boldsymbol R}_{\omega}\hat{\boldsymbol R}_{\omega'}^{T}\ket{\Psi})
\end{equation}
If $\ket{\Psi}$ is a Gaussian pure sate,  the relation
\begin{equation}
\int_{0}^{\infty}d\omega_{1}d\omega_{2}M(\omega,\omega_{1})\Omega(\omega_{1},\omega_{2})M(\omega_{2},\omega')=\frac{1}{4}\Omega(\omega,\omega')
\end{equation}
is satisfied. An incoming momentum operator can be expressed in terms of $\hat{Q}_{\omega}$ and $\hat{P}_{\omega}$ as
\begin{align}
\hat{\Pi}_{in}(v)&=-i\int_{0}^{\infty}\sqrt{\frac{\omega}{4\pi}}\left(\frac{\hat{Q}_{\omega}+i\hat{P}_{\omega}}{\sqrt{2}}e^{-i\omega v}-\frac{\hat{Q}_{\omega}-i\hat{P}_{\omega}}{\sqrt{2}}e^{i\omega v}\right)d\omega \nonumber \\
&=\int_{0}^{\infty}\sqrt{\frac{\omega}{2\pi}}\left(-\sin{(\omega v)}\hat{Q}_{\omega}+\cos{(\omega v)}\hat{P}_{\omega}\right)d\omega
\end{align}
So, the mode $A$ operators are also written by
\begin{align}
\hat{q}_{A}&=\int_{-\infty}^{\infty}q_{A}^{in}(v)\hat{\Pi}_{in}(v)dv=\int_{0}^{\infty}d\omega {\boldsymbol v}_{A}(\omega)^{T}\hat{\boldsymbol R}_{\omega}\\
\hat{p}_{A}&=\int_{-\infty}^{\infty}p_{A}^{in}(v)\hat{\Pi}_{in}(v)dv=\int_{0}^{\infty}d\omega {\boldsymbol u}_{A}(\omega)^{T}\hat{\boldsymbol R}_{\omega}
\end{align}
where
\begin{align}
{\boldsymbol v}_{A}(\omega)&\equiv \int_{-\infty}^{\infty}dv \sqrt{\frac{\omega}{2\pi}}q_{A}(v)(-\sin{(\omega v)},\cos{(\omega v)})^{T} \\
{\boldsymbol u}_{A}(\omega)&\equiv \int_{-\infty}^{\infty}dv \sqrt{\frac{\omega}{2\pi}}p_{A}(v)(-\sin{(\omega v)},\cos{(\omega v)})^{T} 
\end{align}
We make a symplectic transformation 
\begin{align}
\left(
\begin{array}{c}
\hat{Q}_{A} \\
\hat{P}_{A} 
\end{array}\right)\equiv
S_{A}\left(
\begin{array}{c}
\hat{q}_{A} \\
\hat{p}_{A}
\end{array}\right)=\left(
\begin{array}{c}
\int_{0}^{\infty}d\omega{\boldsymbol V}_{A}(\omega)^{T}\hat{{\boldsymbol R}}_{\omega} \\
\int_{0}^{\infty}d\omega{\boldsymbol U}_{A}(\omega)^{T}\hat{{\boldsymbol R}}_{\omega}
\end{array}\right)
\end{align}
such that the operators $(\hat{Q}_{A},\hat{P}_{A})$ are in the standard form where
\begin{align}
\left(
\begin{array}{c}
{\boldsymbol V}_{A}(\omega) \\
{\boldsymbol U}_{A}(\omega)
\end{array}\right)\equiv S_{A}\left(
\begin{array}{c}
{\boldsymbol v}_{A}(\omega) \\
{\boldsymbol u}_{A}(\omega)
\end{array}\right).
\end{align}
Then, its partner mode $B$ can be given by
\begin{align}
\label{QBin}
\hat{Q}_{B}&=\int_{0}^{\infty}d\omega{\boldsymbol V}_{B}^{T}(\omega)\hat{{\boldsymbol R}}_{\omega} \\
\label{PBin}
\hat{P}_{B}&=\int_{0}^{\infty}d\omega{\boldsymbol U}_{B}^{T}(\omega)\hat{{\boldsymbol R}}_{\omega}
\end{align}
where the functions ${\boldsymbol V}_{B}(\omega),\ {\boldsymbol U}_{B}(\omega)$ are determined from the covariance matrix \cite{TYH} as
\begin{align}
\label{partner V}
{\boldsymbol V}_{B}(\omega)&=\frac{\sqrt{1+g^2}}{g}{\boldsymbol V}_{A}(\omega)-\frac{2}{g}\int_{0}^{\infty}d\omega_{1}d\omega_{2}\Omega(\omega,\omega_{1})M(\omega_{1},\omega_{2}){\boldsymbol U}_{A}(\omega_{2})\\
\label{partner U}
{\boldsymbol U}_{B}(\omega)&=-\frac{\sqrt{1+g^2}}{g}{\boldsymbol U}_{A}(\omega)-\frac{2}{g}\int_{0}^{\infty}d\omega_{1}d\omega_{2}\Omega(\omega,\omega_{1})M(\omega_{1},\omega_{2}){\boldsymbol V}_{A}(\omega_{2}).
\end{align}
At this time, the covariance matrix of mode $A$ and $B$ including the correlations between $A$ and $B$ is
\begin{align}M_{AB}&\equiv\left(
\begin{array}{cccc}
\bra{\Psi}\hat{Q}_{A}^{2}\ket{\Psi} & {\rm Re}\bra{\Psi}\hat{Q}_{A}\hat{P}_{A}\ket{\Psi} & \bra{\Psi}\hat{Q}_{A}\hat{Q}_{B}\ket{\Psi} & \bra{\Psi}\hat{Q}_{A}\hat{P}_{B}\ket{\Psi} \\
{\rm Re}\bra{\Psi}\hat{P}_{A}\hat{Q}_{A}\ket{\Psi} & \bra{\Psi}\hat{P}_{A}^{2}\ket{\Psi} & \bra{\Psi}\hat{P}_{A}\hat{Q}_{B}\ket{\Psi} & \bra{\Psi}\hat{P}_{A}\hat{P}_{B}\ket{\Psi} \\
\bra{\Psi}\hat{Q}_{B}\hat{Q}_{A}\ket{\Psi} & \bra{\Psi}\hat{Q}_{B}\hat{P}_{A}\ket{\Psi} & \bra{\Psi}\hat{Q}_{B}^{2}\ket{\Psi} & {\rm Re}\bra{\Psi}\hat{Q}_{B}\hat{P}_{B}\ket{\Psi} \\
\bra{\Psi}\hat{P}_{B}\hat{Q}_{A}\ket{\Psi} &\bra{\Psi}\hat{P}_{B}\hat{P}_{A}\ket{\Psi} &{\rm Re}\bra{\Psi}\hat{P}_{B}\hat{Q}_{B}\ket{\Psi} & \bra{\Psi}\hat{P}_{B}^{2}\ket{\Psi} 
\end{array}\right) \\
&=\left(
\begin{array}{cccc}
\frac{\sqrt{1+g^{2}}}{2} & 0 & \frac{g}{2} & 0 \\
0 & \frac{\sqrt{1+g^{2}}}{2} &0 & -\frac{g}{2} \\
\frac{g}{2} & 0 & \frac{\sqrt{1+g^{2}}}{2} & 0 \\
0 & -\frac{g}{2} & 0 & \frac{\sqrt{1+g^{2}}}{2}
\end{array}\right)
\end{align}

Substituting \eqref{partner V} and \eqref{partner U} for Eqs.\eqref{QBin} and \eqref{PBin}, we can get the formula \eqref{in_formula_Q} and \eqref{in_formula_P} after simple calculation using
 the relation
\begin{align}
\hat{\boldsymbol R}_{\omega}=\sqrt{\frac{2}{\omega \pi}}\int_{-\infty}^{\infty}dv\left(
\begin{array}{c}
-\sin{(\omega v)} \\
\cos{(\omega v)}
\end{array}\right)\hat{\Pi}_{in}(v).
\end{align}
\section{DERIVATION OF EQS.\eqref{out_formula_Qw} AND \eqref{out_formula_Pw}\label{derivation_B}}
For the vacuum Gaussian state $\ket{0_{in}}$ which is defined by $\hat{a}_{\omega}\ket{0_{in}}=0\ (\forall \omega)$, the covariance matrix is simplified as
\begin{align}
M(\omega_{1},\omega_{2})=\frac{1}{2}\left(
\begin{array}{cc}
\delta(\omega_{1}-\omega_{2}) & 0 \\
0 & \delta(\omega_{1}-\omega_{2}) \\
\end{array}\right).
\end{align}
Equations \eqref{partner V} and \eqref{partner U} become
\begin{align}
\label{Vb}
{\boldsymbol V}_{B}(\omega)&=\frac{\sqrt{1+g^2}}{g}{\boldsymbol V}_{A}(\omega)-\frac{1}{g}\Omega {\boldsymbol U}_{A}(\omega) \\
\label{Ub}
{\boldsymbol U}_{B}(\omega)&=\frac{\sqrt{1+g^2}}{g}{\boldsymbol U}_{A}(\omega)-\frac{1}{g}\Omega {\boldsymbol V}_{A}(\omega)
\end{align}
where
\begin{equation}
\Omega\equiv\left(
\begin{array}{cc}
0 & 1 \\
-1 & 0
\end{array}\right).
\end{equation}
Substituting \eqref{Vb} and \eqref{Ub} for Eqs.\eqref{QBin} and \eqref{PBin}, 
\begin{align}
\int_{0}^{\infty}d\omega (\Omega {\boldsymbol U}_{A}(\omega))^{T}\hat{\boldsymbol R}_{\omega}&=\int_{0}^{\infty}d\omega \sqrt{\frac{2}{\pi\omega}}(\Omega {\boldsymbol U}_{A}(\omega))^{T}\left(
\begin{array}{c}
-\sin(\omega v) \\
\cos (\omega v)
\end{array}\right)\hat{\Pi}_{in}(v) \\
&=\frac{1}{\pi}\int_{-\infty}^{\infty}dv\int_{-\infty}^{\infty}dv'\int_{0}^{\infty}d\omega P_{A}^{in}(v')\sin(\omega(v'-v))\hat{\Pi}_{in}(v) \nonumber \\
&=\frac{1}{2\pi i}\int_{-\infty}^{\infty}dv\int_{-\infty}^{\infty}dv'P_{A}^{in}(v')\int_{0}^{\infty}d\omega(e^{-i\omega(v-v')}-e^{i\omega(v-v')})\hat{\Pi}_{in}(v)\nonumber \\
&=-\int_{-\infty}^{\infty}dv\int_{-\infty}^{\infty}dv'\left(\frac{i}{2\pi}\int_{-\infty}^{\infty}d\omega {\rm sgn}(\omega)e^{-i\omega(v-v')}\right)P_{A}^{in}(v')\hat{\Pi}_{in}(v)\nonumber \\
&=-\int_{-\infty}^{\infty}dv\int_{-\infty}^{\infty}dv'\Delta(v-v')P_{A}^{in}(v')\hat{\Pi}_{in}(v)
\end{align}
where we have defined
\begin{equation}
\Delta(v-v')\equiv \frac{i}{2\pi}\int_{-\infty}^{\infty}d\omega\ {\rm sgn}(\omega)e^{-i\omega(v-v')}
\end{equation}
Therefore, we get
\begin{equation}
\hat{Q}_{B}=\int_{-\infty}^{\infty} Q_{B}^{in}(v)\hat{\Pi}_{in}(v)dv,\ \hat{P}_{B}=\int_{-\infty}^{\infty} P_{B}^{in}(v)\hat{\Pi}_{in}(v)dv
\end{equation}
where the weighting functions are given by
\begin{align}
Q_{B}^{in}(v)&=\frac{\sqrt{1+g^2}}{g}Q_{A}^{in}(v)+\frac{1}{g}\int_{-\infty}^{\infty}dv' \Delta(v-v')P_{A}^{in}(v') \\
P_{B}^{in}(v)&=-\frac{\sqrt{1+g^2}}{g}P_{A}^{in}(v)+\frac{1}{g}\int_{-\infty}^{\infty}dv' \Delta(v-v')Q_{A}^{in}(v').
\end{align}
Finally, using $v=p(u)$ and the relation \eqref{Pi_relation}, we can obtain
\begin{equation}
\hat{Q}_{B}=\int_{-\infty}^{\infty}Q_{B}(u)\hat{\Pi}_{out}(u)du,\ \hat{P}_{B}=\int_{-\infty}^{\infty}P_{B}(u)\hat{\Pi}_{out}(u)du
\end{equation}
where $Q_{B}(u)$ and $P_{B}(u)$ are given by Eqs.\eqref{out_formula_Qw} and \eqref{out_formula_Pw}.

\end{document}